\documentclass[prb,aps,showpacs]{revtex4}

\usepackage{epsfig}
\usepackage{dcolumn}
\usepackage{amsmath}
\usepackage{color}
\usepackage{amssymb}
\usepackage {graphicx}

\newcommand{\beq}{\begin{equation}}
\newcommand{\eeq}{\end{equation}}
\newcommand{\bea}{\begin{eqnarray}}
\newcommand{\eea}{\end{eqnarray}}
\newcommand{\be}{\begin{equation}}
\newcommand{\ee}{\end{equation}}

\newcommand{\e}{\varepsilon}

\newcommand{\f}{\frac}

\begin{document}

\title{Spin excitations in layered antiferromagnetic metals and superconductors}
\author {W. Rowe$^{1}$}
\email{wwang@ufl.edu}
\author{J. Knolle$^{2}$}
\author{I. Eremin$^{3}$}
\email{ieremin@tp3.rub.de}
\author{P.J.~Hirschfeld$^{1}$}

\affiliation{$^1$ Department of Physics, University of Florida, Gainesville, USA\\
$^2$Max-Planck-Institut f\"{u}r Physik komplexer Systeme, D-01187
Dresden, Germany\\
$^3$Institut f\"ur Theoretische Physik III,
Ruhr-Universit\"at Bochum, D-44801 Bochum, Germany}

\date{\today}

\begin{abstract}
The proximity of antiferromagnetic
order in high-temperature superconducting materials is considered
a possible clue to the electronic excitations which form superconducting
pairs. Here we study the transverse and longitudinal spin excitation
spectrum in a one-band model in the pure spin density wave (SDW)
state and in the coexistence state of SDW and the superconductivity. We
start from a Stoner insulator and study the evolution of the spectrum with doping,
including distinct situations with only hole pockets, with only electron
pockets and with pockets of both types.  In addition to the usual spin-wave modes,
in the partially gapped cases we find significant weight of low-energy particle-hole excitations.
We  discuss the implications of our findings for neutron scattering experiments and for theories of Cooper-pairing in
the metallic SDW state.
\end{abstract}


\maketitle

\section{Introduction}
\label{sec:1}

Understanding the microscopic origin of interacting and competing phases with
multiple order parameters (OP) is one of the central issues in
condensed matter.
This topic is particularly important for describing the complex
phases of layered transition metal oxides
such as the high-T$_c$ superconductors in which
superconductivity occurs upon hole or electron doping of an antiferromagnetic Mott insulator.
As a consequence of the competition between the Mott insulator and superconductivity,
a rich variety of ground states
may  emerge, such as charge, spin, or current density wave states,
all of which have been proposed to play a role in the pseudogap region of the
high-T$_c$ phase diagram.
For \emph{n}-type cuprates, there is growing experimental evidence for the
evolution of superconductivity in the
background of the commensurate spin density wave (AF) state with ordering momentum {\bf Q}$=(\pi,\pi)$ \cite{review-greene}.

One of the most interesting questions concerns the evolution of the spin excitations in a pure AF state upon introducing carriers by doping,
and the possible coexistence of such a state with $d_{x^2-y^2}$-wave superconductivity. The former question was studied in detail within a strong-coupling
models such as the $t-J$ model Hamiltonian\cite{inui,siggia,belinicher,sushkov}. It starts from a model of localized spins at zero doping which interact via the antiferromagnetic superexchange interaction and then introduces  doped holes on the AF background.  This approach is well justified in the hole-doped cuprates which are Mott insulators in the AF state. At the same time, experimental studies of the moderately electron-doped cuprates\cite{review-greene} and parent iron-based superconductors\cite{hirschfeld-review} show that both remain metallic in the AF state. In addition, a recent DMFT study of the undoped electron-doped cuprates with $T^{\prime}$ crystal structure \cite{weber} suggests that their insulating behavior is more due to
the presence of magnetic long-range order than to  charge transfer gap physics.  This implies that a conventional weak-coupling description of the spin excitations may be applied for the $n$-doped case. This approach was initially developed in the nineties\cite{Schrieffer1989}, and more recently shown to describe well the spin wave spectra of the parent cuprate superconductor La$_2$CuO$_4$\cite{coldea,peres} and iron-based superconductors\cite{knolle}. At the same time, less is known about the spin excitation spectra in the doped situation when the AF state in the electron-doped cuprates is still metallic.

In this paper, we employ the RPA formalism within a single-band Hubbard %
framework to describe the evolution of the spin excitations
from a Stoner insulator to the antiferromagnetic metal. In particular, we analyze the evolution of the transverse and longitudinal spin excitations when the system experiences a transition from
(i) the Stoner insulator to the antiferromagnetic metal with electron Fermi surface pockets centered at $(\pm \pi,0)$ and $(0,\pm \pi)$ points of the Brillouin Zone, and (ii) the Lifshitz transition between two types of the AF metal when in addition to the electron pockets the second small hole Fermi surface pockets appear at $(\pm \pi/2, \pm \pi/2)$ points of the Brillouin zone (BZ).  Furthermore, motivated by the observed coexistence of the AF and
$d_{x^2-y^2}$-wave superconductivity in the electron-doped cuprates we compute the spin excitations in the coexistence region. We find that the Goldstone mode in the transverse channel remains gapless in the coexistence regime and that the excitations in the transverse channel are dominated by the renormalized spectrum of the spin waves. At the same time, we find that the excitations
in the longitudinal channel include a resonance mode at the commensurate momentum close to $(\pi,\pi)$. The simultaneous coexistence of the longitudinal
resonance and the transverse spin waves 
opens up an interesting possibility to use inelastic neutron scattering (INS) to identify the microscopic coexistence of superconductivity and antiferromagnetism. We also stress the importance of well-defined particle-hole branches of the spin excitation spectrum which should also be observable in INS and should help to confirm our general picture.

Our starting point for investigating the coexistence of
superconductivity and SDW
 order is the Hamiltonian
\begin{eqnarray}
\mathcal{H} & = & \sum_{\bf k\sigma} \varepsilon_{\bf k}
c^{\dagger}_{\bf k \sigma} c_{\bf k \sigma} + \sum_{\bf
k,k^\prime,\sigma} U c_{{\bf k}\sigma}^{\dagger} c_{{\bf k+Q}\sigma}
c_{{\bf k^\prime+Q} \bar{\sigma}}^{\dagger} c_{{\bf
k^\prime}\bar{\sigma}}\nonumber
\\&&+\sum_{{\bf k,p,q,\sigma}}V_{\bf
q}\, c_{{\bf k+q}\sigma}^{\dag}c_{{\bf
p-q}\bar{\sigma}}^{\dag}c_{{\bf p}\sigma}c_{{\bf k}\bar{\sigma}}
\label{eq:1}
\end{eqnarray}
where $c_{\bf k\sigma}^{\dag}$ ($c_{\bf k\sigma}$) creates
(annihilates) an electron with spin $\sigma$ and  momentum ${\bf
k}$. We consider a two-dimensional system with  normal state
tight-binding energy dispersion $\varepsilon_{\bf k}=-2t\left( \cos
k_x + \cos k_y \right) + 4 t^\prime \cos k_x \cos k_y - \mu$ and hopping
matrix elements between nearest ($t$) and next-nearest ($t'$) neighbors. The chemical potential $\mu$ controls the doping, $x$ which is
determined by $n=1+x$.
The second and third terms in Eq.~(\ref{eq:1})  lead in the appropriate
Hartree-Fock factorization to the
emergence of commensurate SDW order and superconductivity,
respectively.
While it is generally assumed that both phases emerge
from the same underlying interaction, its renormalization due to
vertex corrections gives rise to a different effective interaction
in each channel, thus justifying the Hamiltonian of Eq.(\ref{eq:1}).
\cite{Schrieffer1989}
\begin{figure}[!t]
\includegraphics[width=1.0\linewidth]{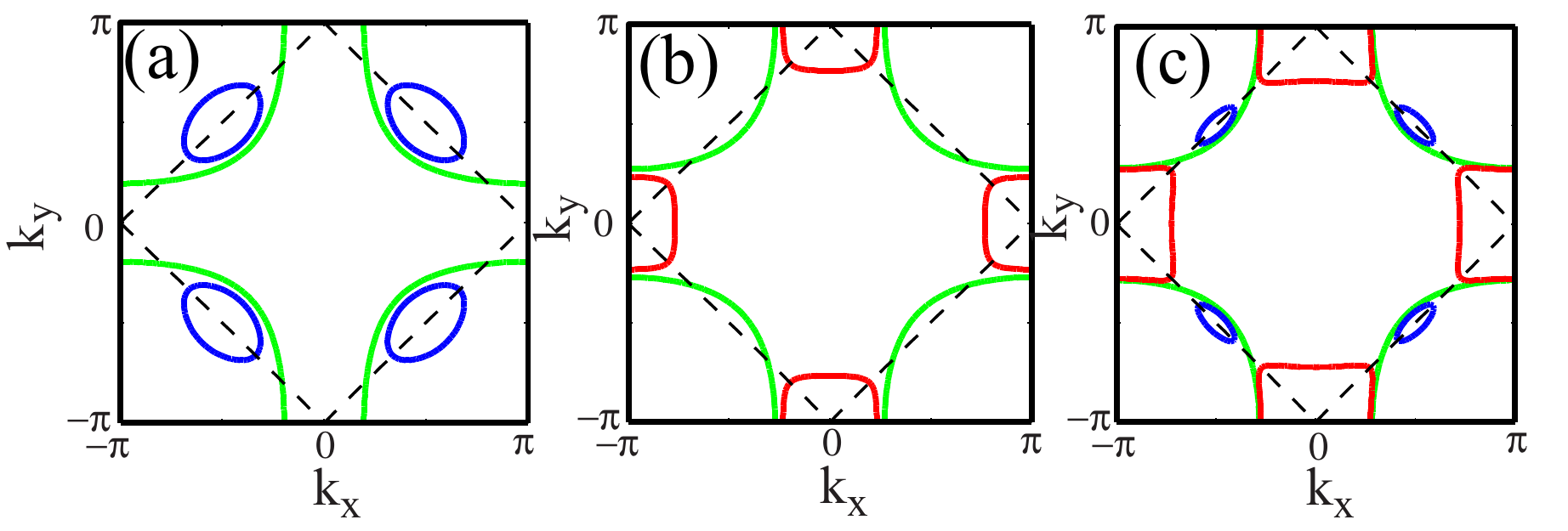}
\caption{color online) Three possible types of  Fermi surface topology in the commensurate SDW state in layered cuprates either for hole (a) or electron doping (b)  and (c) analyzed in this paper. Due to the mean-field SDW order the original large
Fermi surface (green curve) folds, yielding small hole pockets (blue curves) centered around $(\pm \pi/2, \pm \pi/2)$ and
electron pockets (red curves) centered around $(\pm \pi, 0)$ [($(0, \pm \pi)$)] points of the BZ. For larger doping and smaller sizes of the AF gap
both types of the pockets can be present. As argued in the text, the commensurate SDW order becomes unstable once the hole pockets appear
around $(\pm \pi/2, \pm \pi/2)$ points of the BZ. This occurs due to overall negative spin stiffness of the commensurate spin waves in this case, see Fig.\ref{fig3}(a).} \label{fig1}
\end{figure}

We first focus on the pure SDW ordered state, and
consider the case $V_{\bf q}=0$ in Eq. (\ref{eq:1}).
After decoupling the second term
via a mean-field (MF) approximation and diagonalizing the
resulting MF Hamiltonian via a conventional SDW Bogoliubov
transformation, we obtain two electronic bands (labeled $\alpha$ and
$\beta$) in the reduced Brillouin zone (RBZ) with dispersions
$E_{\bf k}^{{\alpha,\beta}} = \e^+_{\bf k}  \pm
\sqrt{\left(\e^-_{\bf k}\right)^{2}+W^{2}}$
where  $W= U/2 \sum_{{\bf k^\prime},\sigma}
 \langle c^{\dagger}_{\bf k^\prime+Q, \sigma} c_{\bf
k^\prime, \sigma} \rangle {\rm sgn}\sigma$ is the SDW order
parameter which is determined self-consistently for a given $U$, and
$\varepsilon^\pm_{\bf k}=\left( \varepsilon_{\bf k} \pm
\varepsilon_{\bf k+Q} \right)/2$. For completeness, one also has to include the self-consistent determination of the chemical potential. Depending on the strength of the SDW gap and the doping level, the system can be either a (i) Stoner insulator for the half-filled case and large value of $W$; or  (ii) SDW metal with small size electron and hole Fermi surface pockets for intermediate values of $W$ and non-zero electron or hole doping, respectively. Typical  Fermi surfaces in the SDW metal are shown in Fig.~\ref{fig1}.

The dynamical spin susceptibility for the longitudinal, $zz$, and
the transverse, $+-$, components is defined as
\begin{eqnarray}
\chi^{lm}({\bf q}, {\bf q}^{\prime},\Omega) &  = &  \int dt \left[\frac{i}{2N}\langle T S^{l}_{\bf q} (t) S^{m}_{\bf -q}(0)\rangle\right] e^{i\Omega t}
\end{eqnarray}
where $lm=zz,+-$. As mentioned above, the antiferromagnetic ordering at {\bf Q}$ =(\pi,\pi)$ doubles
the unit cell and requires  accounting for the  breaking of  translational symmetry\cite{Schrieffer1989,Frenkel1992}. As a result, the
total susceptibility in the transverse channel is a $2\times 2$ matrix
\begin{eqnarray}
 \hat \chi^{+-} &\hskip -0.2cm = \hskip -0.1cm &
\begin{pmatrix}
\chi^{+-}({\bf q},{\bf q},\Omega) & \chi ^{+-}({\bf q},{\bf q+Q},\Omega)  \\
\chi ^{+-}({\bf q+Q},{\bf q},\Omega) & \chi ^{+-}({\bf q+Q},{\bf q+Q},\Omega)
\end{pmatrix}~
\end{eqnarray}
Then, within RPA the susceptibility is obtained via solving a Dyson equation and the result can be expressed\cite{Schrieffer1989} as
\begin{equation}
\hat{\chi}_{RPA}^{+-}=\left(\hat{1}-\hat{U}\hat{\chi}_{0}^{+-}\right)^{-1}\cdot \hat{\chi}_{0}^{+-},
\end{equation}
and the bare components are given by
\begin{eqnarray}
\chi^{+-}_0({\bf q},{\bf q},\Omega) &=& - \frac{1}{2}{\sum_{{\bf k},\gamma}}^\prime \left(
1+\frac{\varepsilon^-_{\bf k} \varepsilon^-_{\bf k+q} - W^2} {\sqrt{ \left( \varepsilon^-_{\bf k} \right)^{2}+
W^{2}} \sqrt{\left(\varepsilon^-_{\bf k+q} \right)^{2}+
W^2}}  \right) \frac{f(E^{\gamma}_{\bf k+q})-f(E^{\gamma}_{\bf
k})}{\Omega +i0^+ -E^{\gamma}_{\bf k+q}+E^{\gamma}_{\bf k}} \nonumber \\
&& \hspace{-1.5cm} - \frac{1}{2}\sum_{{\bf k},\gamma\neq \gamma^{\prime}}^\prime \left(1-\frac{\varepsilon^-_{\bf k}
\varepsilon^-_{\bf k+q}-W^2}
{\sqrt{\left(\varepsilon^-_{\bf k}\right)^{2}+ W^{2}}
\sqrt{\left(\varepsilon^-_{\bf k+q}\right)^{2}+ W^{2}}}\right)
\frac{f(E^{\gamma^{\prime}}_{\bf k+q})-f(E^{\gamma}_{\bf
k})}{\Omega+i0^+-E^{\gamma^{\prime}}_{\bf k+q} +E^{\gamma}_{\bf k}} \nonumber \\
\label{chi_0_qq}
\end{eqnarray}
with $\gamma=\alpha,\beta$
and for the Umklapp term
\begin{eqnarray}
\lefteqn{\chi^{+-}_0({\bf q},{\bf q+Q},\Omega) = } && \nonumber\\
&&  \frac{W}{2} {\sum_{\bf k}}^\prime \left(
\frac{1}{\sqrt{ \left( \varepsilon^-_{\bf k+q} \right)^{2}+
W^{2}}} - \frac{1}{\sqrt{\left(\varepsilon^-_{\bf k} \right)^{2}+
W^2}}  \right)
\left( \frac{f(E^{\alpha}_{\bf k+q})-f(E^{\alpha}_{\bf
k})}{\Omega +i0^+ -E^{\alpha}_{\bf k+q}+E^{\alpha}_{\bf k}}- \frac{f(E^{\beta}_{\bf
k+q})-f(E^{\beta}_{\bf k})}{\Omega+i0^+ -E^{\beta}_{\bf k+q}+ E^{\beta}_{\bf
k}}\right) \nonumber \\
&&  - \left(
\frac{1}{\sqrt{ \left( \varepsilon^-_{\bf k+q} \right)^{2}+
W^{2}}} + \frac{1}{\sqrt{\left(\varepsilon^-_{\bf k} \right)^{2}+
W^2}}  \right)
\left( \frac{f(E^{\beta}_{\bf k+q})-f(E^{\alpha}_{\bf
k})}{\Omega +i0^+ -E^{\beta}_{\bf k+q}+E^{\alpha}_{\bf k}}- \frac{f(E^{\alpha}_{\bf
k+q})-f(E^{\beta}_{\bf k})}{\Omega+i0^+ -E^{\alpha}_{\bf k+q}+ E^{\beta}_{\bf
k}}\right)  \nonumber \\
\label{chi_0_qqpQ}
\end{eqnarray}
where $f(\epsilon)$ is the Fermi function and the prime refers to the sum over the magnetic (reduced) Brillouine Zone (MBZ). Observe that we have $\chi^{+-}_{0}({\bf q,q+Q}) =\chi^{+-}_{0}({\bf q+Q,q})$ and the expression for  $\chi^{+-}_{0}({\bf q+ Q},{\bf q+ Q},\Omega)$ can be obtained from  Eq.
(\ref{chi_0_qq}).

The Umklapp component is absent for the longitudinal susceptibility and the RPA expression is given by
\begin{equation}
\chi^{zz}_{RPA}({\bf q},{\bf q},\Omega) = \frac{\chi^{zz}_{0}({\bf q},{\bf q},\Omega)}{1-U\chi^{zz}_{0}({\bf q},{\bf q},\Omega)}
\end{equation}
with the bare longitudinal susceptibility in the form
\begin{eqnarray}
\chi^{zz}_0({\bf q},{\bf q},\Omega) &\hskip -.2cm=\hskip -0.1cm& - {\frac{1}{2}\sum_{{\bf k},\gamma}}^\prime \left(\hskip -0.1cm
1+\frac{\varepsilon^-_{\bf k} \varepsilon^-_{\bf k+q} + W^2} {\sqrt{ \left( \varepsilon^-_{\bf k} \right)^{2}+
W^{2}} \sqrt{\left(\varepsilon^-_{\bf k+q} \right)^{2}+
W^2}} \hskip -0.05cm \right)   \frac{f(E^{\gamma}_{\bf k+q})-f(E^{\gamma}_{\bf
k})}{\Omega +i0^+ -E^{\gamma}_{\bf k+q}+E^{\gamma}_{\bf k}} \nonumber \\
&& \hspace{-1.9cm} - \frac{1}{2}\sum_{{\bf k},\gamma\neq \gamma^{\prime}}^\prime \left(1-\frac{\varepsilon^-_{\bf k}
\varepsilon^-_{\bf k+q}+W^2}
{\sqrt{\left(\varepsilon^-_{\bf k}\right)^{2}+ W^{2}}
\sqrt{\left(\varepsilon^-_{\bf k+q}\right)^{2}+ W^{2}}}\right)
\frac{f(E^{\gamma}_{\bf k+q})-f(E^{\gamma^{\prime}}_{\bf
k})}{\Omega+i0^+-E^{\gamma}_{\bf k+q} +E^{\gamma^{\prime}}_{\bf k}}
\label{chi_0}
\end{eqnarray}
Here, we note the opposite sign of the $W^2$ term in the coherence factors of the longitudinal and transverse susceptibilities.
\begin{figure}[!t]
\includegraphics[width=1.0\linewidth]{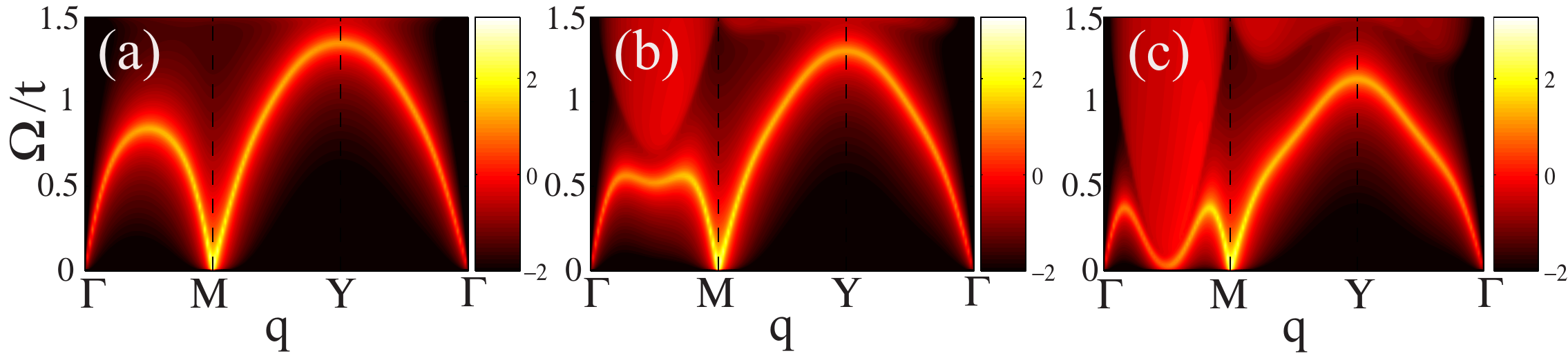}
\caption{(color online) Calculated spin wave dispersion $\Omega$ vs. ${\bf q}$ in units of $\pi/a$ along the symmetry route $(0,0)\to(\pi,\pi)\to(0,\pi)\to(0,0)$ of the BZ for a Stoner insulator at zero doping (x=0), $W=0.75$t and $t^{\prime}=0$ (a), $t^{\prime}=0.2t$ (b), and $t^{\prime}=0.35t$ (c). Here, we fixed the magnitude of $W$ by using $U=1.40t$ and the chemical potential $\mu=0.00t$ (a) $\mu = -0.30t$ (b), and $\mu= -0.68t$ (c), accordingly. The intensity for Im$\chi_{RPA}^{+-}({\bf q},{\bf q},\Omega)$ is shown on the log scale.} \label{fig2}
\end{figure}
Although the general structure of the spin susceptibility in the SDW state at {\bf Q} is known\cite{Schrieffer1989,Frenkel1992},  it is useful to mention its main features. In particular, below T$_N$ spin-rotational invariance is broken and $\chi^{+-}\neq 2\chi^{zz}$.
The imaginary part of the
transverse component is gapless and displays the Goldstone mode at ${\bf Q} = (\pi,\pi)$ and $\Omega \to 0$ for any temperature in the range $0<T<T_N$. The latter is guaranteed by the fact that the condition of the pole formation in the RPA part of the transverse spin susceptibility coincides with the mean-field equation for $W$ and, as clearly seen, is valid for any doping level as soon as the SDW order exists. At the same time, the longitudinal component of the spin susceptibility at {\bf Q} is gapped by twice the SDW gap magnitude, $W$.

What is less well known, however, is the behavior of spin excitations away from {\bf Q}. At half-filling (x=0) and zero temperature, the Fermi surface is gapped for any value of $W$ and $t^{\prime}$ enforced by the
self-consistent calculation of the chemical potential.\cite{remark1}  The excitations in the transverse channel of the Stoner insulator are spin waves - collective spin modes of the antiferromagnetic ground state - with a dispersion all over the BZ\cite{peres}. Due to full gapping of the Fermi surface, the particle-hole Stoner excitations and the spin waves are separated in energy and may interact only around $\Omega_{p-h}({\bf q})$, whose magnitude is controlled by the values of the SDW gap magnitude $W$ and $t^{\prime}/t$ ratio. For $t^{\prime}=0$ the onset of the particle-hole continuum is gapped at least up to $\Omega_{p-h}({\bf Q})=2W$. This is because the top of the lower $\beta$-band and the bottom of the upper $\alpha$-band are located at the magnetic BZ boundary, {\it i.e.} $\cos k_x + \cos k_y = 0$, at energies $-W$ and $+W$, respectively. Therefore, there exists a degenerate manifold of {\bf q} wave vectors for which $ \Omega_{p-h}({\bf q})=2W$. As a result, the spin waves do not interact with the particle-hole continuum for sufficiently large values of $W$ and look identical to those obtained within a Heisenberg model of localized spins which interact via an AF exchange between nearest neighbors, $J_1\sim \frac{t^2}{U}$, see Fig.\ref{fig2}(a).

The situation changes for non-zero values of $t^{\prime}/t$. Its existence (i) introduces non-degenerate positions of the top of the $\beta-$band and bottom of the $\alpha$-band, (ii) reduces the overall magnitude of the gap in the particle-hole continuum and shifts it to lower energies at the $(\pi/2,\pi/2)$ point of the BZ as clearly seen from Fig.\ref{fig2}(b)-(c). To understand its origin, observe that for any non-zero $t^{\prime}/t$ the bottom of the upper $\alpha$-band is
located at $(\pm \pi, 0)$ and $(0, \pm \pi)$  points of the BZ  at energy $-4t^{\prime}+W-\mu > 0$, while the top of the lower $\beta-$band is located at $(\pm \pi/2, \pm \pi/2)$  points of the BZ at energy $-W-\mu < 0$. As a result, the smallest gap between
both bands which determines also the lowest position of the particle-hole continuum  occurs at $\Omega_{p-h}({\bf q}_1) = 2W-4t^{\prime}$ for ${\bf q}_{1}=(\pm \pi/2, \pm \pi/2)$. For increasing $t^{\prime}/t$ ratio and constant value of $W$, the spin waves are bounded from above at momentum ${\bf q}_{1}=(\pm \pi/2, \pm \pi/2)$ and form a local minimum there at energies below $2W-4t^{\prime}>0$.  In particular, in Fig.\ref{fig2}(b) it occurs below 0.7$t$ with $W=0.75t$ and $t^{\prime}=0.2t$ and is shifted to much lower energies for $t^{\prime}=0.35t$ for a fixed $W=0.75t$ as shown in Fig.\ref{fig2}(c). Notice also that for {\it zero doping} we always find either an insulating SDW state (which guarantees the finite gap at  momentum ${\bf q}_{1}$) or a paramagnetic metallic state for $U<U_c$.

The observation of the local minimum for a finite $t^{\prime}$ at ${\bf q}_1$ is due to an interaction of spin waves with the particle-hole continuum. This  is certainly a signature of weak-coupling as it requires $2W$ to be of the same order as $4t^{\prime}$. This effect would not occur for the localized $J_1-J_2$ model, where $J_2$ refers to the antiferromagnetic exchange between the next-nearest neighbors. The inclusion of $J_2$  usually only lowers the position of the maximum of the spin wave dispersion at the $Y$ point of the BZ, an effect clearly reproduced in the weak-coupling calculations as well [compare Fig.\ref{fig2}(a) and (c)]. At the same time, within the localized model the particle-hole excitations  always remain gapped by the large value of $U$ [$W$]. Correspondingly the local minimum in the spin susceptibility at  ${\bf q}_1$ never forms in this case.

To finish the discussion of the spin waves for the undoped case, we further notice that the  spin wave dispersion obtained is symmetric with respect to the  $(0,0)$ and $(\pi,\pi)$ points, which reflects the fact that both
are equivalent symmetry points of the magnetic BZ. At the same time, the absolute intensity of the spin waves is different and is determined by the SDW matrix elements which are suppressed around the $\Gamma$-point. Indeed, from Eq.(\ref{chi_0_qq}) one finds that at low $\Omega$ the non-vanishing contribution to the intensity comes from the interband ($\alpha \to \beta$ and visa versa) transitions  which are proportional to the SDW coherence factor
$c_{\bf k,q}^{inter} = \left(1-\frac{\varepsilon^-_{\bf k}
\varepsilon^-_{\bf k+q}-W^2}
{\sqrt{\left(\varepsilon^-_{\bf k}\right)^{2}+ W^{2}}
\sqrt{\left(\varepsilon^-_{\bf k+q}\right)^{2}+ W^{2}}}\right)$. For ${\bf q \sim Q}$ one finds $\varepsilon^-_{\bf k+Q} \approx - \varepsilon^-_{\bf k}$ and $c^{inter}_{\bf k,q\sim Q} \sim 2$,  whereas it is $c^{inter}_{\bf k,q \sim 0} \propto \left(1-\frac{(\varepsilon^-_{\bf k})^2-W^2}{(\varepsilon^-_{\bf k})^2+W^2}\right)$ for ${\bf q \sim 0}$.
%
\begin{figure}[!t]
\includegraphics[width=1.0\linewidth]{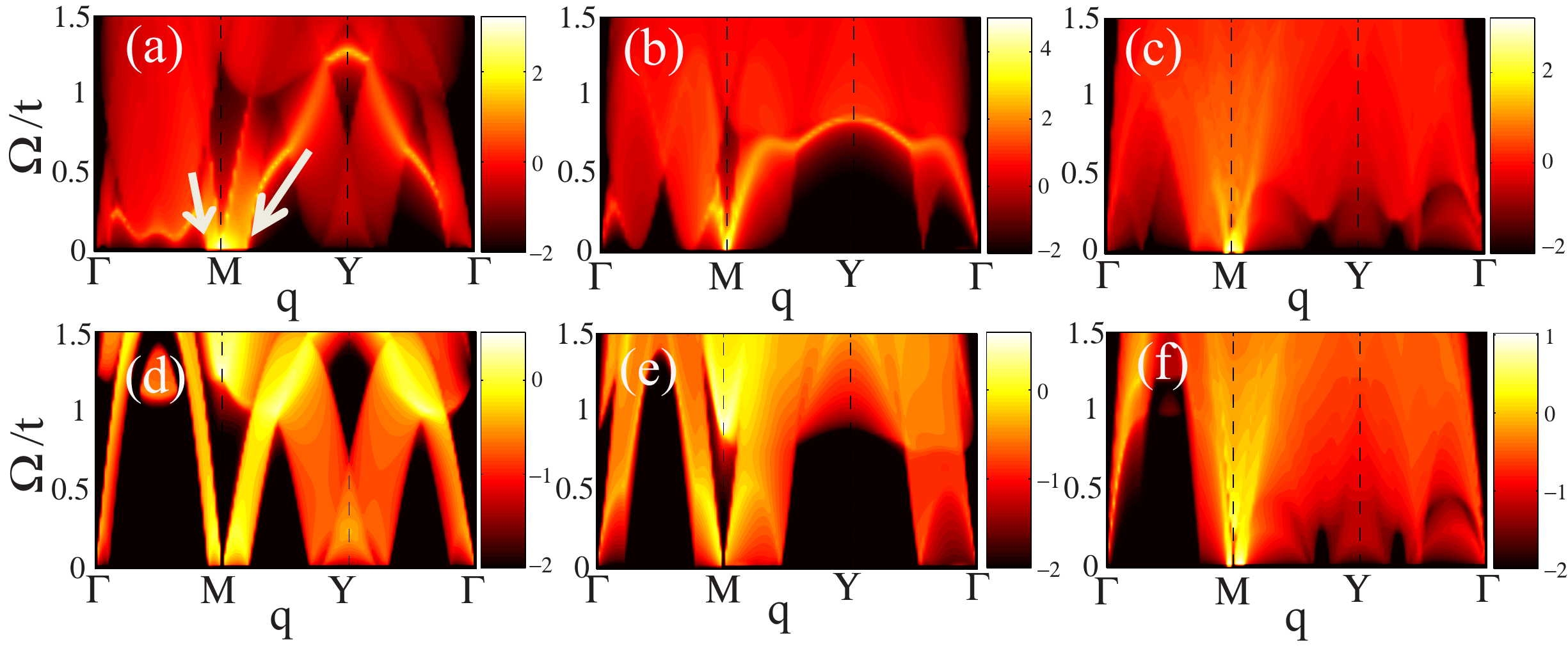}
\caption{color online) Calculated transverse, Im$\chi_{RPA}^{+-}({\bf q},{\bf q},\Omega)$, (upper panel) and longitudinal, Im$\chi_{RPA}^{zz}({\bf q},{\bf q},\Omega)$ (lower panel) spin excitation spectra  $\Omega$ vs. ${\bf q}$ in units of $\pi/a$ for the metallic commensurate SDW state with $t^{\prime}/t = 0.35$ and $U=1.3875t$. (a),(d) refer to the hole doping, x=-0.05, $W=0.61t$, $\mu=-0.8819t$, (b),(e) refer to the  electron doping , $x=0.10$, $W=0.4404t$, $\mu=-0.4284t$, and (c),(f) refer to the electron doping $x=0.14$, $W=0.12t$, $\mu=-0.302t$. The corresponding Fermi surface topology is shown in Fig. \ref{fig1}. The intensity in $states/t$ is shown on a log scale. The white arrows in (a) denote the incommensurate momentum associated with $2k_F$ scattering due to hole FS pockets shown in Fig.\ref{fig1}(a). Observe also the difference in the intensity maps between upper and lower panels.} \label{fig3}
\end{figure}
%

Let us continue by looking at the spin excitations of the commensurate AF order upon electron or hole doping. As we are interested in the metallic ground state, we will only study the situation of the AF metal for finite $t^{\prime}/t$ ratio and remain at very low temperatures. We remind the reader that in this case the Fermi surface topology for the electron- and hole-doped AF metal is different. In the  hole-doped case the chemical potential crosses first the top of the lower $\beta$-band, which is located at $(\pm \pi/2,\pm \pi/2)$ points of the BZ, while for the electron doping  the chemical potential shifts to the bottom of the upper $\alpha-$band which results in the FS electron pockets at $(\pm \pi, 0)$ and $(0, \pm \pi)$.

For the hole-doped case and $t^{\prime}/t>0$, we find that the commensurate antiferromagnetic order is unstable for any finite doping, consistent with
previous analyses\cite{Frenkel1992,Chubukov94}. An instability of the commensurate SDW order can be clearly seen from Fig.\ref{fig3}(a) where we show show an example of the spin excitations obtained for the commensurate SDW order and $x=-0.05$. Note that despite a commensurate SDW order parameter the spin excitations form gapless dispersive {\it incommensurate} modes at ${\bf q}_{i}$ in addition to the strongly damped Goldstone mode which exists at {\bf Q}, despite the fact that commensurate ordering at $(\pi,\pi)$ was assumed in the mean field decoupling. This signals that the commensurate SDW order is not a stable solution.
The origin of this instability is related to the appearance of the small FS hole pockets, whose presence introduces two effects. The first one is related to the spin stiffness of the commensurate spin excitations at {\bf Q}. In particular, in contrast to the undoped case, there is an additional contribution to the spin stiffness which arises due to intraband $\beta-\beta$ transitions which are now gapless. To see this, we expand the dispersion of the lower $\beta-$band for $U>>t$ close to the $(\pm \pi/2,\pm \pi/2)$ points which yields $E_{\bf k}^{\beta}=-\mu-W-\frac{p_{||}^2}{2m_{||}}-\frac{p_{\perp}^2}{2m_{\perp}}$,
where $p_{||}=(k_x-k_y)/2$, $p_{\perp}=(k_x+k_y)/2$ and $m_{||}=(8t^{\prime})^{-1}$, and $m_{\perp}=(16t^2/W-8t^{\prime})^{-1}$. As follows from the analysis of the denominator of the transverse spin susceptibility at ${\bf Q}$, the spin stiffness, $\rho_s$ acquires a finite correction in the doped
SDW metal\cite{Frenkel1992,Chubukov94,Singh91} as $\rho_s=\rho_s^0(1-z)$ where $z=2U\frac{\sqrt{m_{\perp}m_{||}}}{\pi}$ is proportional to the Pauli susceptibility of the $\beta$-band, and $\rho_s^{0}$ is the bare spin stiffness in the undoped case. As one clearly sees, $z>1$ for large $U$ which
indicates that the commensurate SDW order is unstable upon hole doping. On the contrary, for the opposite case $U<<t$ the expansion yields $E_{\bf k}^{\beta}=-\mu-\frac{p_{||}^2}{2m_{||}}-v_{\perp}p_{\perp}+\frac{p_{\perp}^2}{2m_{\perp}}$ where here $m_{\perp}=m_{||}=(8t^{\prime})^{-1}$ and
$v_{\perp} \sim t$. This indicates that for $t^{\prime}<t$ the dispersion along $p_{\perp}$ is essentially linear. As a result the static susceptibility of the $\beta$-band will have singular behavior at $2k_F$. Therefore, even if the Pauli susceptibility of $\beta$ electrons yields $z<1$, there is an additional source of instability associated with $2k_F$ scattering on the hole pockets. To verify which of these two instabilities occurs for $U\sim t$, used in our numerical calculations, we analyzed the behavior of the denominator of the RPA spin susceptibility and found that for the case of Fig.\ref{fig3}(a) we have $z>1$. Therefore, the instability of the commensurate SDW order  for $U \sim t$ and hole doping occurs due to negative corrections to the spin stiffness.

For electron doping the situation is different. In this case the chemical potential moves towards 
the $\alpha-$band which has a minimum at $(\pm \pi,0)$ and $(0, \pm \pi)$ points of the BZ. Expanding now the dispersion of the $\alpha-$band around them yields $E_{\bf k}^{\alpha}=4t^{\prime}-\mu+W+\frac{k_{x}^2}{2m}+\frac{k_{y}^2}{2m}$ where $m=(4t^{\prime})^{-1}$. Observe that this expansion holds for any ratio of $U/t$. As in the case of the hole pockets, there is a contribution of the $\alpha-\alpha$ scattering to the spin stiffness. However, it is easy to see that for the intermediate values of $U \sim t$ and $t^{\prime}=0.35t$ one has $z<1$. In addition, there are additional $k^4$ corrections in the dispersions (proportional to $t^2/W$) which further reduce the value of $z$. We verified it numerically by varying values of $U$ and still finding commensurate SDW order to be a stable solution for  electron doping.

Thus, one of the main results of our paper is that for non-zero $t^{\prime}/t$ ratio there is a strong anisotropy in the
evolution of the commensurate SDW order in the hole- and electron-doped cases [Note that the situation is just opposite for the negative sign of $t^{\prime}$]. While for hole doping the commensurate SDW order is unstable due to an additional response associated with the hole pockets at $(\pm \pi/2, \pm \pi/2)$ which introduce negative spin stiffness to the spin waves, the influence of the electron pockets appearing at $(\pm \pi,0)$ or $(0,\pm \pi)$ is less dramatic and the commensurate SDW persists to a much larger doping. Observe, for example, that for $x=0.1$ (Fig/\ref{fig3}(b)) the spin waves disperse all over the BZ despite the damping introduced by the $\alpha-\alpha$ scattering which is clearly seen in Fig.\ref{fig3}(e).
It is tempting to connect our findings to the experimental observations which show that the commensurate AF order persists up to larger doping concentrations in the electron-doped compounds as compared to the hole-doped ones\cite{review-greene}. Despite the fact that our calculations are done in the weak coupling limit, we believe that the main result still holds in the strong coupling case as well. In particular, the appearance of the Fermi surface pockets upon doping introduces a negative correction $z$ to the spin stiffness. Its exact magnitude can be different in the strong coupling case; however, the asymmetry between the energy dispersions near the bottom of the conduction near $(\pm \pi,0)$ [$(0,\pm \pi)$] points of the BZ and the top of the valence band near $(\pm \pi/2, \pm \pi/2)$ points of the BZ will be similar. Therefore, independent of the details we expect that the trend found in the weak-coupling, {\it i.e.} the relative stability of the commensurate AF order in the electron-doped cuprates as compared to the hole-doped ones, will remain.

A natural question arises: when does the commensurate SDW order becomes unstable for the electron-doped case? Upon increase of the electron doping the pockets at $(\pi,0)$ and $(0,\pi)$ grow in size. However, the Pauli susceptibility in two dimensions is doping independent and consequently the doping does not immediately influence the commensurability of the SDW order.  At the same time, it is natural that for increasing $x$ the SDW gap will decrease. This will reduce the gap between the $\alpha$ and the $\beta$ band in principle, at certain doping level one would expect that both hole and electron pockets may appear at the Fermi level, as shown in Fig.\ref{fig1}(c). Although this situation seems to be supported by ARPES experiments in the electron-doped cuprates, the commensurate SDW order will become unstable in this case.  For the same reason as  in the hole-doped case, the presence of the hole pockets at $(\pi/2,\pi/2)$ immediately introduces critical scattering for the commensurate spin waves. An example of this behavior is shown for the spin excitations in Fig.\ref{fig3}(c),(f) where the commensurate SDW order produces two types of the pockets, see Fig.\ref{fig1}(c). One immediately finds that spin stiffness is again negative and there are incommensurate peaks forming around {\bf Q}. Therefore, the system will either develop incommensurate spiral SDW order or it will readjust the chemical potential such that the hole pockets will disappear with a corresponding change of the electron pocket's size.  There is, however, another interesting possibility which we would like to discuss in the next section.

\section{Spin excitations in the coexistence AF+dSC phase}

As shown above, the presence of electron pockets at $(\pm \pi, 0)$ [$(0, \pm \pi)$] points of the BZ do not
make the commensurate AF order at $(\pi,\pi)$ unstable. This in principle opens an interesting possibility to have a
nodeless $d_{x^2-y^2}$-wave superconducting gap in the electron-doped cuprates which would coexist with AF order. Its features have been discussed previously in a number of mean-field  studies\cite{bob,yan}. Nevertheless, our results indicate that in many cases commensurate AF order becomes unstable with doping due to a negative spin stiffness introduced by the dopants carriers. Therefore, to see whether commensurate AF order can coexist with dSC requires a simultaneous analysis of the spin excitations. In addition, the study of the spin excitations itself in the unconventional superconducting state is interesting on its own due to formation of the spin resonance mode at the  wave vector {\bf Q}$=(\pi,\pi)$ and $\Omega=\Omega_{res}$. Such a feedback effect is quite interesting because its presence allows to confirm the phase structure of the superconducting order parameter even in the situation when thermodynamics reveals an isotropic $s$-wave behavior as in some of the iron-based superconductors\cite{hirschfeld-review}. In the coexistence phase of AF+dSC, the situation is trickier due to the presence of the Goldstone mode at {\bf Q} and the anisotropy of the spin fluctuations in the transverse and the longitudinal channel.

To study the spin dynamics in the coexistence phase we follow Ref.\onlinecite{ismer} and apply the SDW Bogoliubov transformation to
the pairing interaction [third term in Eq.(\ref{eq:1})], and
subsequently perform a mean-field decoupling in the
particle-particle channel, keeping anomalous expectation values
of the form $\langle \alpha^\dagger_{{\bf k}, \uparrow}
\alpha^\dagger_{-{\bf k}, \downarrow} \rangle$ and $\langle
\beta^\dagger_{{\bf k}, \uparrow} \beta^\dagger_{-{\bf k},
\downarrow} \rangle$, and their complex conjugates. As in Ref. \onlinecite{ismer} we further neglect terms of the
form $\langle \alpha^\dagger_{{\bf k}, \uparrow}
\beta^\dagger_{-{\bf k}, \downarrow} \rangle$ due to the FS mismatch between the $\alpha$ and $\beta$
bands. As we show below these terms can be safely ignored once the Umklapp Cooper-pairing terms are absent in the coexistence phase. The resulting mean-field Hamiltonian can be
diagonalized by two independent Bogolyubov transformations, yielding
$\Omega_{{\bf k}}^{\gamma} = \sqrt{\left(E_{{\bf
k}}^{\gamma}\right)^{2}+\left(\Delta_{{\bf k}}^{\gamma}\right)^{2}}$
($\gamma = \alpha, \beta$) as the energy dispersion of the two
bands. The SC gaps $\Delta_{{\bf k}}^{\alpha,\beta}$
are determined self-consistently from two coupled gap equations, derived previously \cite{ismer}.
\begin{eqnarray}
\Delta_{{\bf k}}^{\alpha} & = & -\sum_{{\bf p}\in RBZ}\left[  \f{L_{\bf k,p}}{2E_{{\bf
p}}^{\alpha}} \Delta_{{\bf
p}}^{\alpha} +  \f{M_{\bf k,p}}{2E_{{\bf
p}}^{\beta}} \Delta_{{\bf
p}}^{\beta} \right] \nonumber\\
\Delta_{{\bf k}}^{\beta} & = & -\sum_{{\bf p}\in RBZ}\left[  \f{M_{\bf k,p}}{2E_{{\bf
p}}^{\alpha}} \Delta_{{\bf
p}}^{\alpha} + \f{L_{\bf k,p}}{2E_{{\bf
p}}^{\beta}} \Delta_{{\bf
p}}^{\beta}  \right]
\label{gaps}
\end{eqnarray}
where
$L_{\bf k,p} = \left(
V_{\bf k-p} F^{u,v}_{{\bf k},{\bf p}} - V_{\bf k-p+Q} F^{v,u}_{{\bf
k},{\bf p}} \right)$, $M_{\bf k,p} = \left( V_{\bf k-p}N^{v,u}_{{\bf k},{\bf
p}} - V_{\bf k-p+Q} N^{u,v}_{{\bf k},{\bf p}} \right)$
with $N^{x,y}_{{\bf k},{\bf p}}, F^{x,y}_{{\bf k},{\bf p}},
=u^2_{\bf k}x^2_{\bf p} \pm 2u_{\bf k}v_{\bf k}u_{\bf p}v_{\bf p} +
v^2_{\bf k}y^2_{\bf p}$, $x,y=u,v$, $u_{\bf k}^{2}, v_{\bf k}^{2} =
\frac{1}{2} \left[ 1 \pm \frac{\varepsilon^-_{\bf
k}}{\sqrt{\left(\varepsilon^-_{\bf k}\right)^{2}+W^{2}}} \right]$,
and $u_{\bf k} v_{\bf k}  = \frac{W}{2\sqrt{\left(\varepsilon^-_{\bf
k}\right)^{2}+W^{2}}}$.
For the $d_{x^2-y^2}$-wave pairing potential $V_{\bf k-p} = \frac{V_d}{4} \phi_{\bf k} \phi_{\bf p}$ case where $\phi_{\bf k} =  \cos k_x - \cos k_y $, the SDW coherence factors can be factorized\cite{ismer} and one finds
\begin{equation}
\Delta_{\bf
k}^{\gamma} = \varphi_{\bf k} (\Delta_0^{\gamma}+u_{\bf k}v_{\bf
k}\Delta_1^{\gamma})
\end{equation}
with $\Delta_0^\gamma=
F_0^{\alpha}+F_0^{\beta}$,
$\Delta_1^\alpha=-\Delta_1^\beta=F_1^{\beta}-F_1^{\alpha}$. Here,
$F_0^{\gamma} = \f{V_d}{4} \sum_{{\bf p}} D_{\bf p}^\gamma
\varphi_{\bf p}\f{\Delta_{\bf p}^{\gamma}}{\Omega_{\bf p}^{\gamma}}
\tanh\left(\frac{\Omega_{\bf p}^{\gamma}}{2T}\right)$ and
$F_1^{\gamma} = V_d \sum_{{\bf p}} D_{\bf p}^\gamma \varphi_{\bf p}
u_{\bf p} v_{\bf p} \f{\Delta_{\bf p}^{\gamma}}{\Omega_{\bf
p}^{\gamma}}\tanh\left(\frac{\Omega_{\bf p}^{\gamma}}{2T}\right)$.
Note, $D_{\bf p}^\gamma$ is unity if $|E_{\bf k}^{\gamma}|\leq \hbar \omega_D$, and zero otherwise, with $\hbar \omega_D$ being the Debye frequency.
The SC OP in the $\alpha$ and $\beta$-bands are in phase without additional
line nodes, in contrast to the $s$-wave case\cite{ismer,kulic,overhauser}. Nevertheless, the coexistence of commensurate AF and dSC phases can generate a higher harmonic, $\Delta_1$, in the $d_{x^2-y^2}$-wave gap which has an opposite sign on the electron and hole FS pockets. This harmonic is proportional to the magnitude of the SDW and dSC gaps and arises due to Umklapp Cooper-pairing in terms of original fermions, i.e. averages of the type $\langle c_{{\bf k}, \uparrow} c_{{\bf -k-Q},\downarrow}\rangle$. These averages appear naturally in the coexistence phase as the wavevector {\bf Q} becomes the new reciprocal wave vector of the lattice in the SDW state. At the same time, due to additional breaking of the spin rotational symmetry associated with the SDW transition the Umklapp Cooper-pairing terms formally belong now to the spin-triplet component of the Cooper-pair wave function with $m_z=0$ as was discussed previously by several authors\cite{psaltakis,kyung,varelogiannis}. This indicates that the appearance of the $\Delta_1$ would be then associated with an additional phase transition in the dSC+SDW coexistence phase with a further change of the underlying symmetry of the mean-field Hamiltonian, {\it i.e.} from $SU(2) \times U(1)$ to $SO(5)$ symmetry.

For self-consistency the SC gap equations have to be complemented by the new equation which determines the SDW gap value in the dSC+SDW state
\begin{eqnarray}  \label{sdwdsc}
 W=-\frac{U}{2}   {\sum }_{{\bf k}}^{\prime}  \frac{W}{\sqrt{\left(\varepsilon_{\bf k}^{-}+W^2\right)}}\left[\frac{E^\alpha_{\bf k}}{\Omega^{\alpha}_{\bf k}}\tanh\left(\frac{\Omega^{\alpha}_{\bf k}}{2 k_B T}\right)-\frac{E^\beta_{\bf k}}{\Omega^{\beta}_{\bf k}}\tanh\left(\frac{\Omega^{\beta}_{\bf k}}{2 k_B T}\right)\right]
\end{eqnarray}
and that for the chemical potential, $1+x=\sum^{\prime}_{{\bf k}}\left[ 1-\frac{E^{\alpha}_\textbf{k}}{2\Omega^{\alpha}_\textbf{k}} \tanh(\frac{\Omega^{\alpha }_\textbf{k}} {2T})-\frac{E^{\beta}_\textbf{k}}{2\Omega^{ \beta}_\textbf{k}} \tanh(\frac{\Omega^{ \beta}_\textbf{k}} {2T}) \right]$. Note that the mean-field equations are complemented by the calculations of the total energy
\begin{eqnarray}  
\langle \mathcal{H}\rangle =&\sum^{\prime}_{{\bf k}} &E^\alpha_\textbf{k}-\Omega^\alpha_\textbf{k}+E^\beta_\textbf{k}-\Omega^\beta_\textbf{k}+\frac{\left(\Delta^{\alpha}_\textbf{k}\right)^2}{2\Omega^\alpha_\textbf{k}}\tanh\left[{\frac{\Omega^\alpha_\textbf{k}}{2T}}\right] + \frac{\left(\Delta^{\beta}_\textbf{k}\right)^2}{2\Omega^\beta_\textbf{k}}\tanh\left[{\frac{\Omega^\beta_\textbf{k}}{2T}}\right]\nonumber\\
&& + 2 \Omega^\alpha_\textbf{k} f\left(\Omega_\textbf{k}^\alpha\right)+2 \Omega^\beta_\textbf{k} f\left(\Omega_\textbf{k}^\beta\right)+\frac{W^2}{U}
\end{eqnarray}
to guarantee that their solution refers to its minimum.

We further mention that a comparison of our eigenenergies $\Omega_k^{\gamma}$ with those found in Ref.\onlinecite{kyung} by the diagonalization of the $4 \times 4$ matrix allows us to make a connection between our approach of taking the sequential transformations and the exact diagonalization of the mean-field matrix. In particular, it is straightforward to see that the eigenenergies are the same for $\Delta_1=0$. Thus, the use of sequential unitary transformations, performed by us, is exact as long as the triplet component of the Cooper-pair wave function is absent. For $\Delta_1 \neq 0$ the eigenenergies agree only up to terms of the order $\sim  O(\Delta_1)$ but start to differ for the higher order terms. To recover the same energy spectrum  as in Ref.\onlinecite{kyung} for the coexistence phase one needs to take into account the Cooper-pair terms $\langle \alpha^\dagger_{{\bf k}, \uparrow}
\beta^\dagger_{-{\bf k}, \downarrow} \rangle$ in the AF state. They can be particularly important in the case when the magnetic gap becomes small when the interband Cooper-pairing may be non-negligible. Furthermore, the particular form of the pairing interaction in the momentum space can further modify the structure of the superconducting gap equations in the coexistence state.

The expression for the transverse spin susceptibility in the coexistent state within a RPA remains structurally the same as in the pure AF state, see Eq.(3),  except that the bare susceptibilities contain Cooper-pair creation and annihilation processes in addition to the quasiparticle scattering within one band and between the bands.  The full expression is given in the Appendix. The situation is more complicated for the $zz$ component as there is a mixing of the charge and longitudinal spin susceptibilities at {\bf Q} for a finite doping \cite{Frenkel1992}. Nevertheless, the pole in total RPA susceptibility is still determined by $1-\chi^{zz}_0({\bf q},{\bf q},\Omega)$ which allows us to use Eq.(7) also in the coexistence phase.

\begin{figure}[!t]
\includegraphics[width=0.6\linewidth]{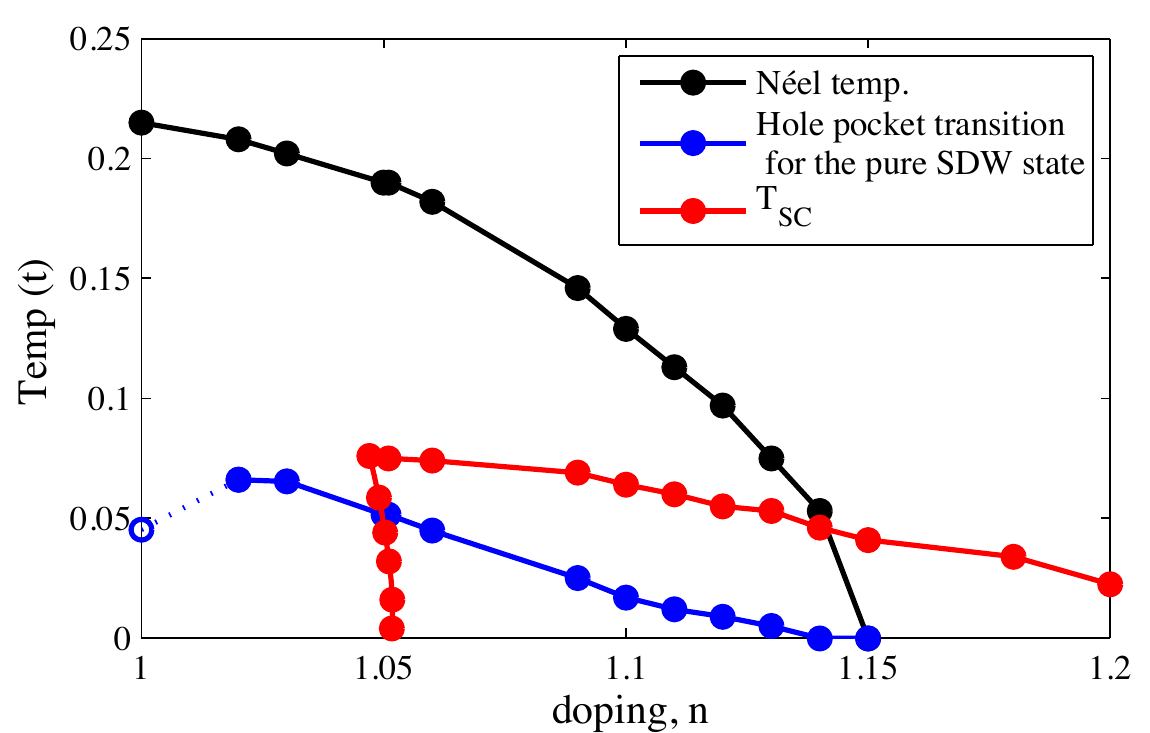}
\caption{color online) Calculated mean-field phase diagram of coexisting $d$-wave superconductivity and commensurate AF order in the electron-doped cuprates for the case $\Delta_1 = 0$. The  solid and dashed blue curve refer to the Lifshitz transition indicating the change of the Fermi surface topology from two type of pockets to the one type of pockets in the pure AF and AF+dSC state, respectively. The open circle for $x=0$ refers to the AF semimetal with two equal sizes electron and hole pockets (high T) to AF insulator (low T) transition.} \label{fig4}
\end{figure}

In Fig.\ref{fig4} we present the results for the mean-field phase diagram of the electron-doped cuprates for coexisting commensurate AF and $d$-wave superconducting order.  Although this type of phase diagram was already obtained in the literature, there are several features which are important to mention. Observe, for example that within a pure SDW phase at finite doping  there is a Lifshitz transition (blue dashed curve) separating phases with a different FS topology with either one or two types of FS pockets. At higher temperatures, both electron and hole type of pockets are present at the FS, while below the Lifshitz transition only the electron pockets are present.  Another interesting feature concerns the character of the phase transition into the coexisting AF+dSC state at low temperatures. In particular, analyzing the free energy we find that the transition from AF to AF+dSC state is of first order as a function of doping, while it becomes second order as a function of temperature.  Furthermore, we notice  that our total energy analysis shows that the stationary solutions for coexistence of SDW and $d-$wave superconductivity in the electron-doped cuprates have always slightly lower free energy in the case $\Delta_1=0$, {\it i.e.} when the triplet component of the Cooper-pairing is absent. In other words in our approach the symmetry of the problem remains $SU(2) \times U(1)$ in the coexistence regime. We believe that it is connected to the fact that we ignored the contribution from the interband Cooper-pair averages $\langle \alpha^\dagger_{{\bf k}, \uparrow}
\beta^\dagger_{-{\bf k}, \downarrow} \rangle$ in the coexistence phase. Although small they could change the balance of the free energy towards the coexistence state  with finite 'triplet' component of the Cooper-pairing. Furthermore, a modification of the momentum dependence of the Cooper-pairing interaction may also change the balance of the mean-field states. This, however, would require a separate analysis which is beyond the scope of the present paper.

Next we compute the transverse spin response in the coexistence region of the phase diagram for three characteristic electron doping levels, $n=1.06$, $n=1.09$, and $n=1.12$ (we fixed T$=0.02t$). First it is easy to check analytically that as soon as $\Delta_{\bf k}^{\alpha}=\Delta_{\bf k}^{\beta}$ which is fulfilled for $\Delta_1=0$ the condition for the pole in transverse susceptibility, $1=U \mbox{Re} \chi_0^{+-}({\bf q=Q},\Omega=0)$, again coincides with the mean-field equation determining the SDW gap  in the coexistence phase, Eq.(\ref{sdwdsc}). Therefore, similar to the pure SDW state the transverse component of the spin susceptibility shows gapless Goldstone mode also in the coexistence region. This is because the pure spin singlet $d-$wave superconductivity alone does not break the spin symmetry of the Hamiltonian. Here, our results do agree with those found very recently in a different formalism Ref.\onlinecite{takimoto}. At the same time, in Ref.\onlinecite{das_new} the spectrum of the transverse spin excitations computed numerically in the coexistence phase for hole doping seems to be gapped at the antiferromagnetic momentum, {\bf Q}, which would contradict our analytical and numerical results. We speculate that this discrepancy arises from the fact that at the hole doping studied in Ref.\onlinecite{das_new} the commensurate AF order is not a stable solution. In particular, although the mean-field solution for the commensurate AF order exists, the resulting spin waves have negative spin stiffness\cite{Frenkel1992} (see also our Fig.3). As a result the true magnetic order in this case is incommensurate.

Although the spin waves are gapless in the coexistence phase, their velocity is affected by the presence of $d-$wave superconductivity.
\begin{figure}[!t]
\includegraphics[width=0.6\linewidth]{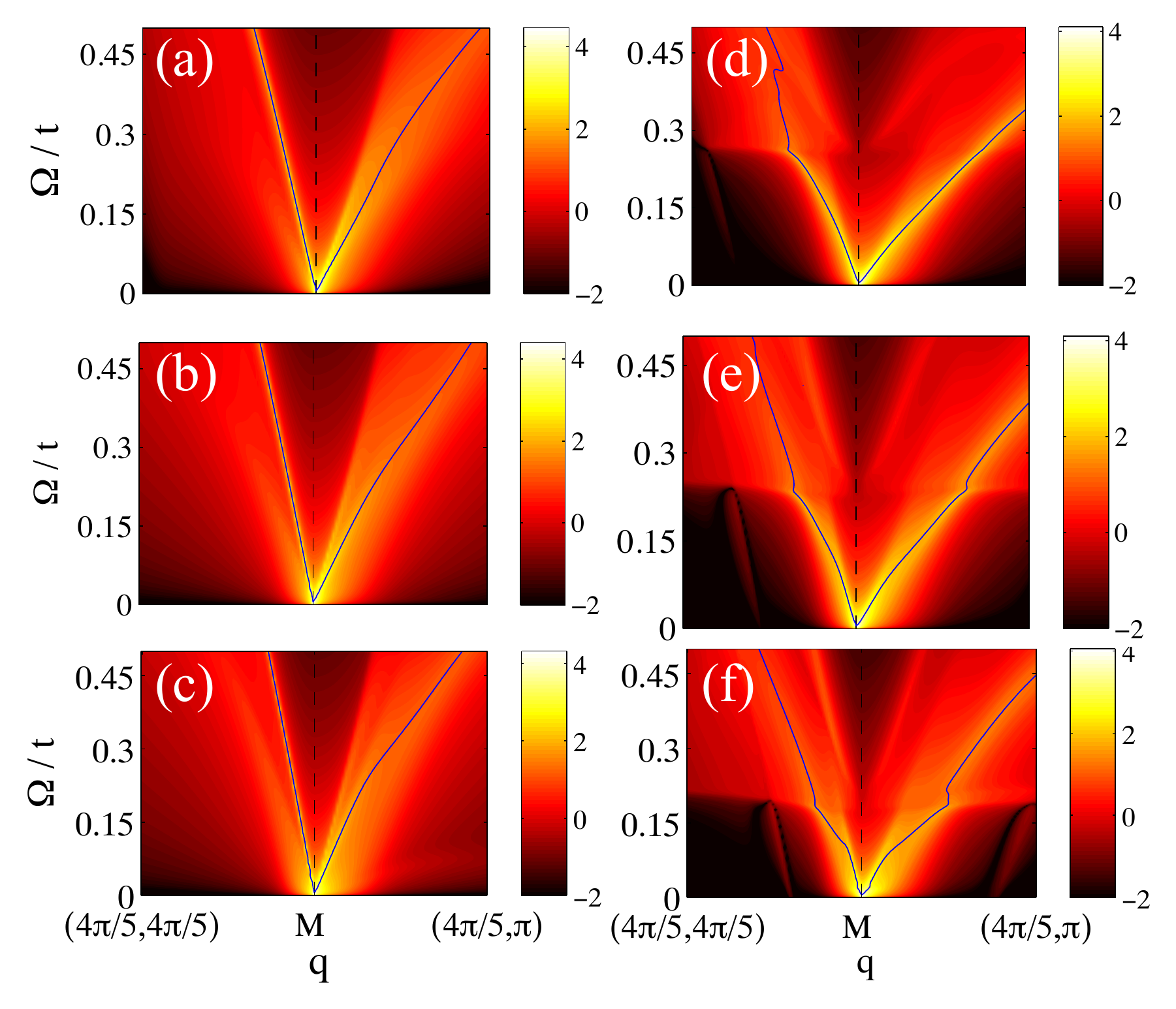}
\caption{color online) Calculated transverse, Im$\chi_{RPA}^{+-}({\bf q},{\bf q},\Omega)$ spin excitation spectra  $\Omega$ vs. ${\bf q}$ in units of $\pi/a$ for three different electron dopings,  $n=1.06$, $n=1.09$, and $n=1.12$ (from upper to lower panel) for the coexisting SDW+dSC state and $\Delta_1=0$ (right panel). For comparison the left panel shows the results for the pure SDW state. The blue lines denote $1=U \mbox{Re} \chi_0^{+-}({\bf q},\Omega)$ condition. The intensity is shown on the log scale. The following parameters are used in the units of $t$  for (a) $\mu=-0.5362$, $W=0.5617$, for (b) $\mu=-0.4573$, $W=0.4617$, for (c) $\mu=-0.3694$, $W=0.3456$, for (d) $\mu=-0.5349$, $W=0.5530$, $\Delta_0=0.0727$, for (e) $\mu=-0.4509$, $W=0.4555$, $\Delta_0=0.0705$, and for (f) $\mu=-0.3624$, $W=0.3458$, $\Delta_0=0.0618$.} \label{fig5}
\end{figure}
In particular, in the Fig.\ref{fig5} (d)-(f) we show the behavior of the transverse, Im$\chi_{RPA}^{+-}({\bf q},{\bf q},\Omega)$ spin excitations for three different electron dopings,  $n=1.06$, $n=1.09$, and $n=1.12$ (from upper to lower panel) in the coexistence state. Observe that in all three cases the spin excitations are gapless but the spin wave velocity is influenced by the superconducting gap. For small energies this can be seen analytically by expanding the
denominator of Eqs. (4) around {\bf q},$\Omega=0$ up to
quadratic order. This procedure leads to the spin wave velocity, $c$  of the form:
\begin{equation}
c^2=\frac{yt^2(1/U-W^2z)}{W^2x^2+(v)(1/U-W^2z)}
\end{equation}
The coefficients for $x$, $y$, $z$ contain in the lowest order  terms proportional  to $\Delta_0^2$. Their exact form is given in the Appendix. The influence of the superconducting gap on the spin wave spectra is also seen  by comparison of the transverse spin excitations spectra of the coexistence SDW+dSc state with that of the pure SDW state which is shown in the left panel, Fig.\ref{fig5}(a)-(c). The effect is particularly strong around $2\Delta_0$ where the spin waves in the coexistence region exhibit a kink structure due to the interaction with the $p-h$ continuum of the intraband $\alpha-\alpha$ and $\beta-\beta$ transitions. They are enhanced due to $d-$wave symmetry of the superconducting gap, {\it i.e.} due to the fact that one finds $\Delta^{\gamma}_{\bf k} = - \Delta^{\gamma}_{\bf k+q_i}$ for incommensurate momenta {\bf q}$_i > (0.8,0.8)\pi$. This then leads to an enhancement of the intraband particle-hole continuum of both bands for $\Omega \approx 2\Delta_0$. As the electron band around $(\pm \pi,0)$ and $(0,\pm \pi)$ points of the BZ always crosses the Fermi level in the SDW state the enhancement of the particle-hole continuum of this band around $2 \Delta_0$ is responsible for the kink structure seen in the spin waves. In other words, the damping effects of the particle-hole continuum on the spin waves are present in both pure metallic SDW and coexisting SDW+dSC states. However, in the coexistence region there is also an effect of the strong renormalization of the spin wave due to the 2$\Delta_0$ structure of the particle-hole continuum of the intraband susceptibility, which then yields the renormalization of the spin wave velocity around $2\Delta_0$. Another interesting feature is that $d-$wave superconductivity stabilizes the commensurate AF state by partial gapping the particle-hole continuum in the coexistence state. Observe, for example that the spin waves computed for $n=1.12$ in the pure SDW state show a tendency towards incommensurability, while in the coexistence state the spin excitations are still commensurate.

\begin{figure}[!t]
\includegraphics[width=0.6\linewidth]{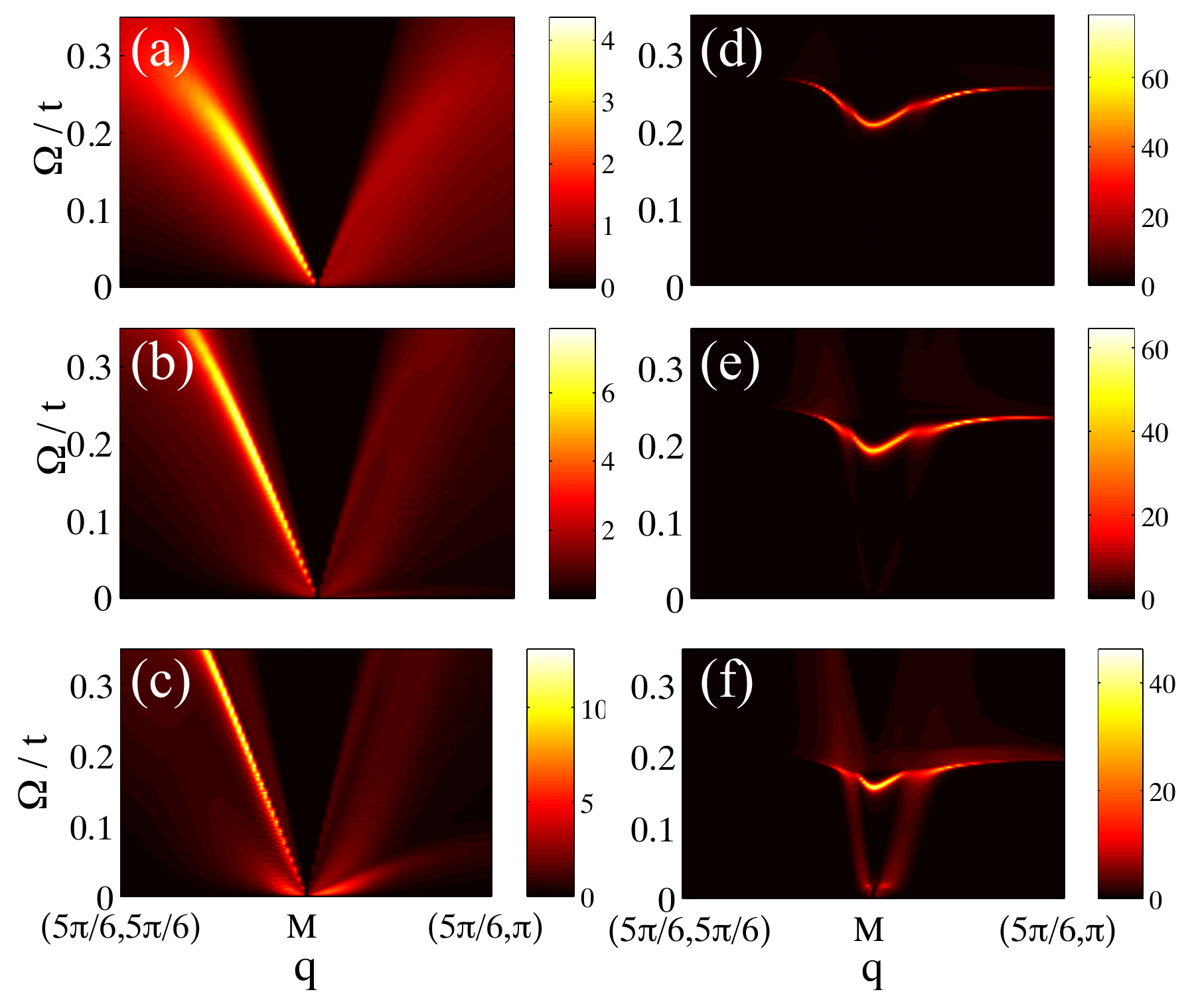}
\caption{color online) Calculated longitudinal, Im$\chi_{RPA}^{zz}({\bf q},{\bf q},\Omega)$ spin excitation spectra  $\Omega$ vs. ${\bf q}$ in units of $\pi/a$ for three different electron dopings,  $n=1.06$, $n=1.09$, and $n=1.12$ (from upper to lower panel) for the coexisting SDW+dSC state and $\Delta_1=0$ (right panel). For comparison the left panel shows the behavior of the longitudinal susceptibility for the pure SDW state. The intensity is shown on the absolute scale.} \label{fig6}
\end{figure}

In Fig.\ref{fig6}(d)-(f) we also show the behavior of the longitudinal response in the coexistence region. These are again compared to the pure SDW state Fig.\ref{fig6}(a)-(c). It is interesting that in this case the same structure of the particle-hole continuum at $2\Delta_0$ for the intraband transitions which generates the strong damping of the spin waves around $2\Delta_0$ in $\chi^{+-}$ introduces the formation of the spin resonance at energies below $2\Delta_0$ in $\chi^{zz}$. In particular, the resonance occurs because of the discontinuous jump of the imaginary part of the intraband susceptibility at $2\Delta_0$. Correspondingly, there is a log singularity of the real part and below it Re $\chi_0({\bf Q},\Omega) \propto \omega^2$ for the intraband susceptibility. As a result, the total Re $\chi_0({\bf Q},\Omega)$ is enhanced at $2\Delta_0$ and the resonance condition $1/U= \mbox{Re}\chi_0({\bf Q},\Omega=\Omega_{res})$ is fulfilled for $\Omega_{res}<2\Delta_0$. This is in contrast to the pure SDW state where the main contribution at {\bf Q} comes from the interband excitations which are gapped by the SDW gap magnitudes. At the same time, the intraband quasiparticle scattering in the SDW state is suppressed for the wavevector ${\bf Q}$. In the coexistence state, in addition to the intraband quasiparticle scattering, there are also intraband Cooper-pairing annihilation processes, which start to contribute to the spin susceptibility at $2\Delta$. These processes are enhanced by the $d-$wave symmetry of the gap and are responsible for the formation of the spin resonance. This resonance, however, now occurs in a very narrow {\bf q} region, which is bounded by the dispersion of the particle-hole continuum. In particular,
the spin resonance in the longitudinal component of the spin susceptibility is bounded from above at the wavevector {\bf Q}  by the onset of the particle-hole continuum associated with breaking of the Cooper-pairs at approximately $2\Delta_0$ and from the left and right at finite $\delta {\bf q}$ by the onset of the continuum associated with intraband quasiparticle scattering which is gapless due to the presence of nodes on the $\alpha$ Fermi surface in the $d-$wave gap. As a result the spin resonance shows a characteristic upward dispersion and then becomes overdamped. This is quite different from the typical downward dispersion of the spin resonance associated with the $d$-wave symmetry of the superconducting state without coexisting long range AF order.

Finally we also note that our expression for the spin susceptibility computed for the metastable state for $\Delta_1\neq 0$ consistently signals its instability, that is we find  $1-U\mbox{Re}\chi^{+-}_0({\bf q=Q},\Omega=0)<0$ in this case. This indicates that the calculations of the spin response in presence of the triplet Cooper-pairing requires one to go beyond the sequential transformations, which we leave for the future studies.

\section{Conclusion}

In this work, we studied  the behavior of the transverse and longitudinal spin excitation spectrum in a one-band
model within a pure spin density wave (SDW) state and in the coexistence region of SDW and d-wave superconductivity. In particular, we analyze the evolution of the transverse and longitudinal spin excitations in  doped AF
metals. We have shown that for the sign of $t^{\prime}$ appropriate for the cuprates,
commensurate spin excitations are stable within weak-coupling only on the electron-doped side of
the phase diagram.  Otherwise,  the results we obtained in the SDW phase have been obtained in various forms by previous authors. However, in addition to the spin wave spectrum in this phase, we have focused on
the evolution of particle hole excitations.  These may play an important role in any instability of such
an AF phase to superconductivity, since the pairing vertex due to exchange of the magnons themselves
is suppressed due to the Adler principle.  Calculations along these lines are in progress.

Furthermore, motivated by the observed coexistence of the AF and $d_{x^2-y^2}$-wave superconductivity in the
electron-doped cuprates we computed the spin excitations in the coexistence region. We find that the Goldstone mode
in the transverse channel remains gapless in the coexistence regime and that the excitations in the transverse channel are dominated by the renormalized spectrum of the spin waves. This happens, however, only if the Umklapp Cooper-
pairing term, $\langle c_{\bf k} c_{\bf -k-Q} \rangle$, which actually belongs to the spin-triplet component of the Cooper-wave function is absent. We show also that within our approach this triplet component never
appears as the free energy is always higher for the solution with non-zero Umklapp terms. This is a subtle point which requires a careful, fully self-consistent treatment of competing order parameters in the problem. At the same time, we find that the excitations in the longitudinal channel include a resonance mode at the commensurate momentum close to
$(\pi,\pi)$. The simultaneous coexistence of the longitudinal resonance and the transverse spin waves opens up an
interesting possibility to use inelastic neutron scattering (INS) to identify the microscopic coexistence of the
superconductivity and antiferromagnetism. We also stress the importance of well-defined particle-hole branches of the spin excitation spectrum which should also be observable in INS and should help to confirm our general picture.

\section{Acknowledgements}

We acknowledge helpful discussions with G. Blumberg, A.V. Chubukov, R. Fernandes, J.-P. Ismer, R. Moessner, D.K. Morr, and J. Schmalian. W. Rowe and P.J.H. were supported by NSF-DMR-1005625. W. Rowe is thankful for hospitality of the Ruhr-University Bochum where the final stage of this work was done.
IE acknowledges financial support of the SFB Transregio 12, Merkur Foundation, and German Academic Exchange Service (DAAD PPP USA No. 50750339) and is thankful to the University of Florida in Gainesville where the project was initiated. JK acknowledges support from the Studienstiftung des deutschen Volkes and the IMPRS Dynamical Processes  in Atoms, Molecules and Solids.

\section{Spin susceptibility in the coexistent AF+dSC state}

Here, we present the expression for the spin susceptibility in the coexistent AF+dSC region. In particular the transverse component
has the form
\bea
\lefteqn{\chi^{+-}_0({\bf q},{\bf q},\Omega) = } &&\nonumber\\
&&{\sum_{{\bf k},\gamma}}^{\prime} \frac{1}{4} \left(
1+\frac{\varepsilon^-_{\bf k} \varepsilon^-_{\bf k+q} - W^2} {\sqrt{ \left( \varepsilon^-_{\bf k} \right)^{2}+
W^{2}} \sqrt{\left(\varepsilon^-_{\bf k+q} \right)^{2}+
W^2}}  \right) \left\{ \left[1+ \frac{E_{\bf k}^{\gamma}E_{\bf k+q}^{\gamma}+\Delta_{\bf k}^{\gamma}\Delta_{\bf k+q}^{\gamma}}{\Omega_{\bf k}^{\gamma}\Omega_{\bf k+q}^{\gamma}}\right] \frac{f(\Omega^{\gamma}_{\bf k+q})-f(\Omega^{\gamma}_{\bf
k})}{\Omega +i0^+ -\Omega^{\gamma}_{\bf k+q}+\Omega^{\gamma}_{\bf k}} \right. \nonumber\\
&& \left. + \frac{1}{2}\left[1- \frac{E_{\bf k}^{\gamma}E_{\bf k+q}^{\gamma}+\Delta_{\bf k}^{\gamma}\Delta_{\bf k+q}^{\gamma}}{\Omega_{\bf k}^{\gamma}\Omega_{\bf k+q}^{\gamma}}\right]
\left( \frac{f(\Omega^{\gamma}_{\bf k+q})+f(\Omega^{\gamma}_{\bf
k})-1}{\Omega +i0^+ +\Omega^{\gamma}_{\bf k+q}+\Omega^{\gamma}_{\bf k}}+ \frac{1-f(\Omega^{\gamma}_{\bf k+q})-f(\Omega^{\gamma}_{\bf
k})}{\Omega +i0^+ -\Omega^{\gamma}_{\bf k+q}-\Omega^{\gamma}_{\bf k}} \right)\right\} \nonumber\\
&& {\sum_{{\bf k},\gamma \neq \gamma^{\prime}}}^{\prime} \frac{1}{4} \left(
1-\frac{\varepsilon^-_{\bf k} \varepsilon^-_{\bf k+q} - W^2} {\sqrt{ \left( \varepsilon^-_{\bf k} \right)^{2}+
W^{2}} \sqrt{\left(\varepsilon^-_{\bf k+q} \right)^{2}+
W^2}}  \right) \left\{ \left[1+ \frac{E_{\bf k}^{\gamma}E_{\bf k+q}^{\gamma^{\prime}}+\Delta_{\bf k}^{\gamma}\Delta_{\bf k+q}^{\gamma^{\prime}}}{\Omega_{\bf k}^{\gamma}\Omega_{\bf k+q}^{\gamma^{\prime}}}\right] \frac{f(\Omega^{\gamma^{\prime}}_{\bf k+q})-f(\Omega^{\gamma}_{\bf
k})}{\Omega +i0^+ +\Omega^{\gamma^{\prime}}_{\bf k+q}-\Omega^{\gamma}_{\bf k}} \right. \nonumber\\
&& \left. + \frac{1}{2}\left[1- \frac{E_{\bf k}^{\gamma}E_{\bf k+q}^{\gamma^{\prime}}+\Delta_{\bf k}^{\gamma}\Delta_{\bf k+q}^{\gamma^{\prime}}}{\Omega_{\bf k}^{\gamma}\Omega_{\bf k+q}^{\gamma^{\prime}}}\right]
\left( \frac{f(\Omega^{\gamma^{\prime}}_{\bf k+q})+f(\Omega^{\gamma}_{\bf
k})-1}{\Omega +i0^+ +\Omega^{\gamma^{\prime}}_{\bf k+q}+\Omega^{\gamma}_{\bf k}}+ \frac{1-f(\Omega^{\gamma^{\prime}}_{\bf k+q})-f(\Omega^{\gamma}_{\bf
k})}{\Omega +i0^+ -\Omega^{\gamma^{\prime}}_{\bf k+q}-\Omega^{\gamma}_{\bf k}} \right)\right\}
\eea
and
\begin{eqnarray}
\lefteqn{\chi^{+-}_0({\bf q},{\bf q+Q},\Omega) = } && \nonumber\\
&&  \frac{W}{4} {\sum_{{\bf k},\gamma}}^\prime \left(
\frac{1}{\sqrt{ \left( \varepsilon^-_{\bf k+q} \right)^{2}+
W^{2}}} - \frac{1}{\sqrt{\left(\varepsilon^-_{\bf k} \right)^{2}+
W^2}}  \right) \left\{
\pm \left(\frac{E_{\bf k+q}^{\gamma}}{\Omega_{\bf k+q}^{\gamma}}+\frac{E_{\bf k}^{\gamma}}{\Omega_{\bf k}^{\gamma}}\right) \frac{f(\Omega^{\gamma}_{\bf k+q})-f(\Omega^{\gamma}_{\bf
k})}{\Omega +i0^+ -\Omega^{\gamma}_{\bf k+q}+\Omega^{\gamma}_{\bf k}} \right.\nonumber \\
&& \pm \left. \left(\frac{E_{\bf k+q}^{\gamma}}{\Omega_{\bf k+q}^{\gamma}}-\frac{E_{\bf k}^{\gamma}}{\Omega_{\bf k}^{\gamma}}\right) \left(\frac{1-f(\Omega^{\gamma}_{\bf k+q})-f(\Omega^{\gamma}_{\bf
k})}{\Omega +i0^+ -\Omega^{\gamma}_{\bf k+q}-\Omega^{\gamma}_{\bf k}} + \frac{f(\Omega^{\gamma}_{\bf k+q})+f(\Omega^{\gamma}_{\bf
k})-1}{\Omega +i0^+ +\Omega^{\gamma}_{\bf k+q}+\Omega^{\gamma}_{\bf k}}\right) \right\} \nonumber \\
&&  + \frac{W}{4} {\sum_{{\bf k},\gamma \neq \gamma^{\prime}}}^\prime \left(
\frac{1}{\sqrt{ \left( \varepsilon^-_{\bf k+q} \right)^{2}+
W^{2}}} + \frac{1}{\sqrt{\left(\varepsilon^-_{\bf k} \right)^{2}+
W^2}}  \right) \left\{ \pm \left(\frac{E_{\bf k+q}^{\gamma}}{\Omega_{\bf k+q}^{\gamma}}+\frac{E_{\bf k}^{\gamma^{\prime}}}{\Omega_{\bf k}^{\gamma^{\prime}}}\right)
 \frac{f(\Omega^{\gamma}_{\bf k+q})-f(\Omega^{\gamma^{\prime}}_{\bf
k})}{\Omega +i0^+ -\Omega^{\gamma}_{\bf k+q}+\Omega^{\gamma^{\prime}}_{\bf k}}  \right. \nonumber \\
&&
\left. \pm \left(\frac{E_{\bf k+q}^{\gamma}}{\Omega_{\bf k+q}^{\gamma}}-\frac{E_{\bf k}^{\gamma^{\prime}}}{\Omega_{\bf k}^{\gamma^{\prime}}}\right)
\left( \frac{1-f(\Omega^{\gamma}_{\bf k+q})-f(\Omega^{\gamma^{\prime}}_{\bf
k})}{\Omega +i0^+ -\Omega^{\gamma}_{\bf k+q}- \Omega^{\gamma^{\prime}}_{\bf k}} + \frac{f(\Omega^{\gamma}_{\bf k+q})+f(\Omega^{\gamma^{\prime}}_{\bf
k})-1}{\Omega +i0^+ +\Omega^{\gamma}_{\bf k+q}+ \Omega^{\gamma^{\prime}}_{\bf k}} \right) \right\} \quad.
\label{chi_0_qqpQsc}
\end{eqnarray}
Here, $+$ or $-$ refers to the $\gamma=\alpha$ and $\gamma=\beta$, respectively. Correspondingly, the longitudinal component has the form
\bea
\lefteqn{\chi^{zz}_0({\bf q},{\bf q},\Omega) = } &&\nonumber\\
&&{\sum_{{\bf k},\gamma}}^{\prime} \frac{1}{4} \left(
1+\frac{\varepsilon^-_{\bf k} \varepsilon^-_{\bf k+q} + W^2} {\sqrt{ \left( \varepsilon^-_{\bf k} \right)^{2}+
W^{2}} \sqrt{\left(\varepsilon^-_{\bf k+q} \right)^{2}+
W^2}}  \right) \left\{ \left[1+ \frac{E_{\bf k}^{\gamma}E_{\bf k+q}^{\gamma}+\Delta_{\bf k}^{\gamma}\Delta_{\bf k+q}^{\gamma}}{\Omega_{\bf k}^{\gamma}\Omega_{\bf k+q}^{\gamma}}\right] \frac{f(\Omega^{\gamma}_{\bf k+q})-f(\Omega^{\gamma}_{\bf
k})}{\Omega +i0^+ -\Omega^{\gamma}_{\bf k+q}+\Omega^{\gamma}_{\bf k}} \right. \nonumber\\
&& \left. + \frac{1}{2}\left[1- \frac{E_{\bf k}^{\gamma}E_{\bf k+q}^{\gamma}+\Delta_{\bf k}^{\gamma}\Delta_{\bf k+q}^{\gamma}}{\Omega_{\bf k}^{\gamma}\Omega_{\bf k+q}^{\gamma}}\right]
\left( \frac{f(\Omega^{\gamma}_{\bf k+q})+f(\Omega^{\gamma}_{\bf
k})-1}{\Omega +i0^+ +\Omega^{\gamma}_{\bf k+q}+\Omega^{\gamma}_{\bf k}}+ \frac{1-f(\Omega^{\gamma}_{\bf k+q})-f(\Omega^{\gamma}_{\bf
k})}{\Omega +i0^+ -\Omega^{\gamma}_{\bf k+q}-\Omega^{\gamma}_{\bf k}} \right)\right\} \nonumber\\
&&{\sum_{{\bf k},\gamma \neq \gamma^{\prime}}}^{\prime} \frac{1}{4} \left(
1-\frac{\varepsilon^-_{\bf k} \varepsilon^-_{\bf k+q} + W^2} {\sqrt{ \left( \varepsilon^-_{\bf k} \right)^{2}+
W^{2}} \sqrt{\left(\varepsilon^-_{\bf k+q} \right)^{2}+
W^2}}  \right) \left\{ \left[1+ \frac{E_{\bf k}^{\gamma}E_{\bf k+q}^{\gamma^{\prime}}+\Delta_{\bf k}^{\gamma}\Delta_{\bf k+q}^{\gamma^{\prime}}}{\Omega_{\bf k}^{\gamma}\Omega_{\bf k+q}^{\gamma^{\prime}}}\right] \frac{f(\Omega^{\gamma^{\prime}}_{\bf k+q})-f(\Omega^{\gamma}_{\bf
k})}{\Omega +i0^+ -\Omega^{\gamma^{\prime}}_{\bf k+q}+\Omega^{\gamma}_{\bf k}} \right. \nonumber\\
&& \left. + \frac{1}{2}\left[1- \frac{E_{\bf k}^{\gamma}E_{\bf k+q}^{\gamma^{\prime}}+\Delta_{\bf k}^{\gamma}\Delta_{\bf k+q}^{\gamma^{\prime}}}{\Omega_{\bf k}^{\gamma}\Omega_{\bf k+q}^{\gamma^{\prime}}}\right]
\left( \frac{f(\Omega^{\gamma^{\prime}}_{\bf k+q})+f(\Omega^{\gamma}_{\bf
k})-1}{\Omega +i0^+ +\Omega^{\gamma^{\prime}}_{\bf k+q}+\Omega^{\gamma}_{\bf k}}+ \frac{1-f(\Omega^{\gamma^{\prime}}_{\bf k+q})-f(\Omega^{\gamma}_{\bf
k})}{\Omega +i0^+ -\Omega^{\gamma^{\prime}}_{\bf k+q}-\Omega^{\gamma}_{\bf k}} \right)\right\}
\eea

To evaluate the spin wave velocity, we expand the
denominator of Eqs. (4) around {\bf q},$\Omega=0$ up to
quadratic order. This procedure leads to the spin wave velocity, $c$  of the form
\begin{equation}
c^2=\frac{yt^2(1/U-W^2z)}{W^2x^2+(v)(1/U-W^2z)}
\end{equation}
where
\begin{equation}
x={\sum_{\bf k}}^{\prime}\frac{1}{\sqrt{\left(\varepsilon^{-}_{\bf k}\right)^2+W^2}\left(\Omega_\textbf{k}^\alpha+\Omega_\textbf{k}^\beta\right)^2}\left(\frac{E_\textbf{k}^\alpha}{\Omega_\textbf{k}^\alpha}-\frac{E_\textbf{k}^\beta}{\Omega_\textbf{k}^\beta}\right)
\end{equation}
\begin{equation}
v={\sum_{\bf k}}^{\prime}\frac{1}{\left(\Omega_\textbf{k}^\alpha+\Omega_\textbf{k}^\beta\right)^3}\left(1-\frac{E_\textbf{k}^\alpha E_\textbf{k}^\beta -\Delta^2}{\Omega_\textbf{k}^\alpha \Omega_\textbf{k}^\beta}\right)
\end{equation}
\begin{equation}
z={\sum_{\bf k}}^{\prime}\frac{1}{\left(\left(\varepsilon^{-}_{\bf k}\right)^2+W^2\right)\left(\Omega_\textbf{k}^\alpha+\Omega_\textbf{k}^\beta\right)}\left(1-\frac{E_\textbf{k}^\alpha E_\textbf{k}^\beta +\Delta^2_{\bf k}}{\Omega_\textbf{k}^\alpha \Omega_\textbf{k}^\beta}\right)
\end{equation}
The coefficient $y$ is comprised of  two terms. The first one arises from the intraband contribution
\begin{eqnarray}
y_1 =  {\sum_{{\bf k} \gamma=\alpha, \beta}}^{\prime} &&\frac{\Delta_{\bf k}^2}{2 \left(\Omega_\textbf{k}^{\gamma}\right)^3}   \left[  \frac{W^2\sin^2k_x}{\left(\left(\varepsilon^{-}_{\bf k}\right)^2+W^2\right)^2}-\frac{\left(\cos k_x+\cos k_y\right)^2 }{\left(\varepsilon^{-}_{\bf k}\right)^2+W^2}\right] \nonumber \\
- &&   \frac{3\varepsilon^{-}_{\bf k}\Delta_{\bf k}^2 E_\textbf{k}^\gamma}{\left(\left(\varepsilon^{-}_{\bf k}\right)^2+W^2\right)\left(\Omega^{\gamma}_{\bf k}\right)^5}   \left(2 \sin^2k_x \cos k_y \pm \frac{\varepsilon_{\bf k}^- \sin^2 k_x}{\sqrt{\left(\varepsilon^{-}_{\bf k}\right)^2+W^2}}\right)
\end{eqnarray}
and the other one from the interband contributions
\begin{eqnarray}
\lefteqn{y_2= \frac{1}{4} {\sum_{{\bf k}, \gamma\neq \gamma^{\prime}}}^{\prime}   \frac{2W^2 \sin^2k_x}{\left(\left(\varepsilon^{-}_{\bf k}\right)^2+W^2\right)^2\left(\Omega^\gamma_{\bf k}+\Omega^{\gamma^{\prime}}_{\bf k}\right)}\left( 1-\frac{E_\textbf{k}^\gamma E_\textbf{k}^{\gamma^{\prime}}-\Delta_{\bf k}^2}{\Omega^{\gamma}_{\bf k} \Omega^{\gamma^{\prime}}_{\bf k}}\right)} && \nonumber \\
&&-\frac{2}{\Omega^{\gamma}_{\bf k} \Omega^{\gamma^{\prime}}_{\bf k}\left(\Omega^{\gamma}_{\bf k} +\Omega^{\gamma^{\prime}}_{\bf k}\right)}\left[ \left( \frac{t^\prime}{t}\cos (k_x+k_y) \pm \frac{ \left(\cos^2 k_x+\cos k_x \cos k_y\right)}{\sqrt{\left(\varepsilon_{\bf k}^{-}\right)^2+W^2}} \mp \frac{W^2 \sin^2 k_x}{\left(\left(\varepsilon_{\bf k}^{-}\right)^2+W^2\right)^{3/2}}  \right) E_\textbf{k}^{\gamma^{\prime}} +\frac{\Delta^2_k}{2t^2} \right] \nonumber \\
&&-\frac{1}{\left(\Omega^{\gamma}_{\bf k}\right)^3 \Omega^{\gamma^{\prime}}_{\bf k}\left(\Omega^{\gamma}_{\bf k} +\Omega^{\gamma^{\prime}}_{\bf k}\right)}   \left(\left(E_\textbf{k}^{\prime\gamma}\right)^2
E_\textbf{k}^{\gamma}E_\textbf{k}^{\gamma^{\prime}}  -2\Delta_{\bf k}^2 \Delta_0^2 \frac{\sin^2 k_x}{t^2} \right)
+\frac{\left(E_\textbf{k}^{\gamma }E_\textbf{k}^{\gamma^{\prime}}+\Delta^2_{\bf k}\right)}{\left(\Omega^{\gamma}_{\bf k}\right)^5 \Omega^{\gamma^{\prime}}_{\bf k}\left(\Omega^{\gamma}_{\bf k} +\Omega^{\gamma^{\prime}}_{\bf k}\right)}   \left(\left(E_\textbf{k}^{\prime \gamma}\right)^2
\left(E_\textbf{k}^{\gamma}\right)^2  +2\Delta_{\bf k}^2 \Delta_0^2 \frac{\sin^2 k_x}{t^2} \right)\nonumber \\
&&-\frac{1}{\left(\Omega^{\gamma}_{\bf k}\right)^2 \left(\Omega^{\gamma}_{\bf k} +\Omega^{\gamma^{\prime}}_{\bf k}\right)^3}\left( 1- \frac{E_\textbf{k}^{\gamma}E_\textbf{k}^{\gamma^{\prime}}-\Delta^2_{\bf k}}{\Omega^{\gamma}_{\bf k} \Omega^{\gamma^{\prime}}_{\bf k}} \right)\left(\left(E_\textbf{k}^{\prime\gamma}\right)^2 \left(E_\textbf{k}^{\gamma}\right)^2 +2\Delta_{\bf k}^2 \Delta_0^2 \frac{\sin^2 k_x}{t^2}\right) \nonumber \\
&&-\frac{1}{\left(\Omega^{\gamma}_{\bf k}\right)^2 \Omega^{\gamma^{\prime}}_{\bf k}\left(\Omega^{\gamma}_{\bf k} +\Omega^{\gamma^{\prime}}_{\bf k}\right)^2}  \left(\left(E_\textbf{k}^{\prime\gamma}\right)^2
E_\textbf{k}^{\gamma}E_\textbf{k}^{\gamma^{\prime}}  -2\Delta_{\bf k}^2 \Delta_0^2 \frac{\sin^2 k_x}{t^2} \right)
+\frac{\left(E_\textbf{k}^{\gamma}E_\textbf{k}^{\gamma^{\prime}}-\Delta^2_{\bf k}\right)}{\left(\Omega^{\gamma}_{\bf k}\right)^4 \Omega^{\gamma^{\prime}}_{\bf k}\left(\Omega^{\gamma}_{\bf k} +\Omega^{\gamma^{\prime}}_{\bf k}\right)^2}   \left(\left(E_\textbf{k}^{\prime \gamma}\right)^2
\left(E_\textbf{k}^{\gamma}\right)^2 +2\Delta_{\bf k}^2 \Delta_0^2 \frac{\sin^2 k_x }{t^2} \right) \nonumber \\
&&-\left[ \frac{\left(E_\textbf{k}^{\gamma}E_\textbf{k}^{\gamma^{\prime}}+\Delta^2_{\bf k}\right)}{\left(\Omega^{\gamma}_{\bf k}\right)^2 \Omega^{\gamma^{\prime}}_{\bf k}\left(\Omega^{\gamma}_{\bf k} +\Omega^{\gamma^{\prime}}_{\bf k}\right)} -\left( 1- \frac{E_\textbf{k}^{\gamma}E_\textbf{k}^{\gamma^{\prime} }-\Delta^2_{\bf k}}{\Omega^{\gamma}_{\bf k} \Omega^{\gamma^{\prime}}_{\bf k}}  \right) \frac{1}{\left(\Omega^{\gamma}_{\bf k} + \Omega^{\gamma^{\prime}}_{\bf k}\right)^2}
\right] \nonumber\\
&&\times \left[ \frac{\Delta^2_{\bf k}\left(E_\textbf{k}^{\prime \gamma}\right)^2+\left(E_\textbf{k}^{\gamma}\right)^2\Delta_0^2 \sin^2 k_x / t^2}{\left(\Omega^{\gamma}_{\bf k}\right)^3} -\frac{4}{\Omega^{\gamma}_{\bf k} } \left[ \left( \frac{t^\prime}{t}\cos (k_x+k_y) \pm\frac{ \left(\cos^2 k_x+\cos k_x \cos k_y\right)}{\sqrt{\left(\varepsilon_{\bf k}^{-}\right)^2+W^2}} \mp\frac{W^2 \sin^2 k_x}{\left(\left(\varepsilon_{\bf k}^{-}\right)^2+W^2\right)^{3/2}}  \right) E_\textbf{k}^{\gamma} +\frac{\Delta^2_{\bf k}}{t^2} \right] \right] \nonumber \\
\end{eqnarray}
Here $ \left(E_\textbf{k}^{\prime \alpha,\beta}\right)^2=8\left[ 2\sin^2 (k_x+k_y)\mp\frac{4 \varepsilon_{\bf k}^- \sin^2 k_x \cos k_y}{\sqrt{\left(\varepsilon_{\bf k}^{-}\right)^2+W^2}}+\frac{ \varepsilon_{\bf k}^{-}\sin^2 k_x}{\left(\varepsilon_{\bf k}^{-}\right)^2+W^2}\right]$ and we further use $\Delta_{\bf k}=\Delta_{\bf k}^{\alpha}=\Delta_{\bf k}^{\beta}$.
\end{appendix}


\begin{thebibliography}{99}
\bibitem{remark1} Note that for $t^{\prime}\neq 0$ there is a solution for SDW gap at finite temperature such that it yields a compensated SDW metal with equal volume Fermi surface pockets of the electron and hole character.
\end{thebibliography}
\end{document}